\documentclass[12pt,]{iopart}
\usepackage{iopams}  

\usepackage{bm,mathrsfs,color}
\usepackage{graphicx}
\usepackage{braket}
\usepackage{color}

\newcommand{\eqref}[1]{(\ref{#1})}

\begin{document}

\title[Weak anti-localization in spin-orbit coupled lattice systems]{
Weak anti-localization in spin-orbit coupled lattice systems}

\author{Hiroshi Hayasaka and Yuki Fuseya}

\address{Department of Engineering Science, University of Electro-Communications, Chofu, Tokyo 182-8585, Japan}
\ead{hayasaka@kookai.pc.uec.ac.jp}
\vspace{10pt}

\begin{abstract}
The quantum correction to electrical conductivity is studied on the basis of two-dimensional Wolff Hamiltonian, which is an effective model for a spin-orbit coupled (SOC) lattice system. It is shown that weak anti-localization (WAL) arises in SOC lattices, although its mechanism and properties are different from the conventional WAL in normal metals with SOC impurities. The interband SOC effect induces the contribution from the interband singlet Cooperon, which plays a crucial role for WAL in the SOC lattice.
It is also shown that there is a crossover from WAL to weak localization in SOC lattices when the Fermi energy or band gap changes. The implications of the present results to Bi--Sb alloys and PbTe under pressure are discussed.
\end{abstract}

%
%
%
%
%

\section{Introduction}
\begin{figure}
 \begin{center}
  \includegraphics[width=8cm]{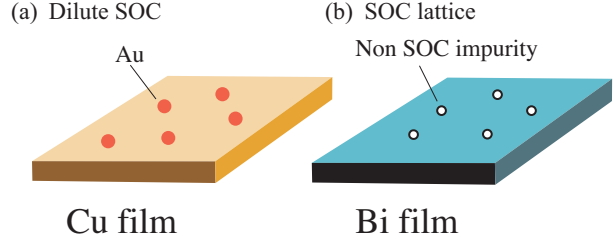}
 \end{center}
 \label{fig:schimatic}
 \caption{Schematic of (a) a dilute SOC system and (b) a SOC lattice system.}
 \end{figure}

It is well known that impurity scattering with spin-orbit coupling (SOC) causes weak anti-localization (WAL) in a two-dimensional system, as shown by Hikami--Larkin--Nagaoka (HLN) \cite{HLN1980}. The spin-relaxation length can be evaluated by analyzing the magnetic field dependence of quantum correction to the electric conductivity, $\delta \sigma (B)$, by using the formula obtained by HLN. The conventional WAL theory considers nearly free electrons in metals with spin-orbit coupled (SOC) impurities that does not conserve electron-spins, e.g., Cu film with Au impurities [Fig. \ref{fig:schimatic} (a)]. This situation is called ``dilute SOC" system in this study. 

Recently, the evaluation of the spin-relaxation length using the HLN formula has become a very important topic in the field of spintronics and topological materials sciences \cite{Grbi2008,Assaf2013,Deorani2014,Peres2014}. However, most target materials in these fields possess strong SOC in the atoms that constitute the lattice, and not the impurities, e.g., Bi film with non SOC impurities [Fig. \ref{fig:schimatic} (b)]. This situation is called ``SOC lattice," which is the opposite limit from the dilute SOC system, in this study. The HLN formula has been used in many SOC lattice systems assuming that the SOC lattice can be described within the same framework as dilute SOC systems \cite{Assaf2013,Deorani2014,Chen2011,Hirahara2014}. However, the application of HLN formula for SOC lattice has not been sufficiently validated, because it was derived for dilute SOC systems.  Naively, there are two possibilities for SOC lattices: (i) WAL arises in SOC lattices as in dilute SOC because the system possess a SOC; or, (ii) WL arises in SOC lattices because they are the opposite limit of dilute SOC. It is not obvious which is correct. This topic of the relationship between dilute SOC and SOC lattices have not been addressed before, although it is similar to the problem of Kondo effect, where both the dilute Kondo and Kondo lattice exhibit minimum resistance \cite{Stewart1984}.

This study aimed to investigate quantum correction to the conductivity in a two-dimensional SOC lattice and examine whether conventional WAL arises. We employed the Wolff Hamiltonian, which is essentially equivalent to the Dirac Hamiltonian, as the minimal model of the SOC lattice \cite{Wolff1964,Fuseya2015}. The quantum correction was calculated using a standard technique of weak localization (WL) and WAL considering the particle--particle ladder-type diagram, i.e., the correction from the Cooperon instability \cite{HLN1980,LuHaiZhou2011,ShanWenYu2012}.	
It is revealed that the SOC lattice exhibits WAL--WL crossover with respect to a function of the ratio of the Fermi energy $E_F$ to the band gap $\Delta$. In the WAL region, there are two essential differences between the SOC lattice and the dilute SOC system. In the dilute SOC system, there is a minimum of the quantum correction $\delta \sigma_{\rm HLN}(B)$ as a function of a magnetic field $B$. In contrast, there is no minimum in the quantum correction of the Wolff model $\delta \sigma_{\rm W}(B)$ in the SOC lattice. Furthermore, it is also shown that WAL arises in the SOC lattice even without spin-relaxation scattering, whereas spin-relaxation scattering is crucial for conventional WAL in dilute SOC systems. The interband effect of SOC and the interband singlet Cooperon instability are essential to this unconventional WAL. Finally, the implications of the present results to Bi--Sb alloys and PbTe under pressure are discussed.

\section{Quantum correction in the Wolff model $\delta \sigma_{\rm W}$}
The Wolff Hamiltonian is an effective model of a system with a strong SOC \cite{Wolff1964,Fuseya2015}. 
We consider the Wolff Hamiltonian with an impurity potential $V(\bm{r})$ in the form
\begin{eqnarray}
\label{hami}
{\cal H}={\cal H}_{\rm W}+V(\bm{r}),
\end{eqnarray}
\begin{eqnarray}
{\cal H}_{\rm W}=\left[
      \begin{array}{cc}
      \Delta  & i\gamma\bm{\sigma} \cdot \bm{k}\\
      -i\gamma\bm{\sigma} \cdot \bm{k} &-\Delta\\
      \end{array}
\right]
\end{eqnarray}
\begin{eqnarray}
V(\bm{r})=u_0\sum_{i}\delta(\bm{r}-\bm{R}_{i}),
\end{eqnarray}
where $\bm{k}=(k_x,k_y)$ is the wave vector, $2\Delta$ is the band gap, $\gamma$ is the velocity, and $\bm{\sigma}$ is the Pauli matrices. $u_0$ is the strength of scattering, and $\bm{R}_i$ is the position of the $i$-th impurity. 
In this study, we assume an isotropic dispersion for simplicity. 

The present $V(\bm{r})$ scatters electrons without changing their bands and spins. However, electrons move by changing their bands and spins because of the SOC in ${\cal H}_{\rm w}$. This effect of interband SOC is a characteristic of SOC lattices.

The momentum scattering rate is given by 
\begin{eqnarray}
\frac{\hbar}{2\tau}&=\pi\sum_{\bm{k}'}\bra{}\bra{\bm{k}}V(\bm{r})\ket{\bm{k}'}\ket{^2}_{\rm imp}\delta(E_{F}-E_{\bm{k}'})\nonumber\\
&=\pi \rho_{0}n_{i}u_{0}^2\frac{\lambda^2+1}{2\lambda^2},
\end{eqnarray}
under the second-order Born approximation.
$\rho_{0}=\Delta/\pi\hbar^2\gamma^2$ is the density of state, $n_i$ is the impurity density per unit volume, and $\lambda=E_F/\Delta$. 
(The absolute values of $\bm{k}$ and $\bm{k}'$ are replaced by the Fermi wavenumber assuming that the quantity on the Fermi surface is relevant.)
We consider a scattering process that does not change the band and spin of the eigenfunction of eq. (2) and discard the scattering process that does. This approximation is essentially the same as that used in Refs. {\cite{LuHaiZhou2011, ShanWenYu2012}}.

The quantum correction to conductivity $\delta\sigma_{\rm{W}}(L)$ is given by \cite{PALeeRev1985,Efros_book1985,FukuyamaRev1985}
\begin{eqnarray}
\label{kubo}
\delta\sigma_{\rm{W}}&=\frac{e^2\hbar}{\pi}\sum_{\bm{q}}\Gamma(\bm{q})\sum_{\bm{k}}v_{\bm{k}}^xv_{\bm{q-k}}^xG^{\rm R}_{\bm{k}}G^{\rm A}_{\bm{k}}G^{\rm R}_{\bm{q-k}}G^{\rm A}_{\bm{q-k}},\nonumber\\
\end{eqnarray}
where W denotes Wolff, and $G^{{\rm R}/{\rm A}}=(E_F-E_{\bm{k}}\pm i\hbar/2\tau)^{-1}$ is the retarded/advanced Green's function, 
$\Gamma(\bm{q})$ is a particle--particle ladder-type diagram that gives the Cooperon, and $v_{\bm{k}}=\gamma(\sqrt{E_{\bm{k}}^2-\Delta^2}/E_{\bm{k}}){\rm cos}\phi_{\bm{k}}$ is the velocity.
$\Gamma(\bm{q})$ satisfies the following Bethe--Salpeter equation
\begin{eqnarray}
\label{BSE}
\Gamma(\bm{q})_{\bm{k}_{\alpha}\bm{k}_{\beta}}=\Gamma_{\bm{k}_{\alpha}\bm{k}_{\beta}}^0+\sum_{\bm{k}_{\mu}}\Gamma_{\bm{k}_{\alpha}\bm{k}_{\mu}}^0G^{\rm R}_{\bm{k}_{\mu}}G^{\rm A}_{\bm{q}-\bm{k}_{\mu}}\Gamma(\bm{q})_{\bm{k}_{\mu}\bm{k}_{\beta}},
\end{eqnarray}
where $\Gamma^{0}$ is given by
\begin{eqnarray}
\Gamma_{\bm{k}_{\alpha}\bm{k}_{\beta}}^0&=\braket{\bra{\bm{k}_{\beta}}V(\bm{r})\ket{\bm{k}_{\alpha}}\bra{-\bm{k}_{\beta}}V(\bm{r})\ket{-\bm{k}_{\alpha}}}_{\rm imp}.
\end{eqnarray}
Here, $\langle\cdots \rangle_{\rm imp}$ denotes the average of the impurities.
The solution of (\ref{BSE}) is of the form,  
\begin{eqnarray}
\label{Cooperon}
\Gamma(\bm{q})_{\bm{k}_{\alpha}\bm{k}_{\beta}}=\frac{\hbar}{2D_{0}\pi \rho_{0}\tau^2}\sum^{2}_{n,m=0}\Gamma_{nm}(\bm{q})e^{i(n\phi_{\alpha}-m\phi_{\beta})},
\end{eqnarray}
where $\Gamma_{00}=\alpha_{\rm t}/(\ell_{\rm t}^{-2}+q^2)$，$\Gamma_{11}=\alpha_{\rm s}/(\ell_{\rm s}^{-2}+q^2)$, $\ell_{\rm t}^{-2}=(\lambda-1)^2/(\lambda+1)^2\alpha_{\rm t}\ell_{0}^{-2}$, and $\ell_{\rm s}^{-2}=2/(\lambda^2-1)\alpha_{\rm s}\ell_{0}^{-2}$. 
$\ell_0=\sqrt{D_0\tau}$ is the mean free path, $D_0=v_{F}^2\tau/2$ is the diffusion constant,
 and $v_{F}=\gamma\sqrt{E_F^2-\Delta^2}/E_{F}$ is the Fermi velocity.
 
$\alpha_{\rm t}$ and $\alpha_{\rm s}$ are key parameters, which determine the magnitude of triplet and singlet Cooperon instabilities, respectively. They are given by 
\begin{eqnarray}
\alpha_{\rm t} =\frac{4}{\lambda^2+3}, \quad
\alpha_{\rm s} =-\frac{(\lambda^2-1)^2}{2(\lambda^2+1)^2}.
\end{eqnarray}
We can neglect  $\Gamma_{22}$ because this term does not provide a divergent contribution. $\Gamma_{nm}$ becomes zero for $n\neq m$ after integration with respect to $\bm{q}$.  
$\Gamma_{nm}$ corresponds to the elements of a multiplet basis representation by representing $\Gamma(\bm{q})_{\bm{k}_{\alpha}\bm{k}_{\beta}}$ in the following form, $(\bra{{\bm{k}_{\beta}}}\otimes\bra{{-\bm{k}_{\beta}}})\Gamma(\ket{{\bm{k}_{\alpha}}}\otimes\ket{{-\bm{k}_{\alpha}}})$ for $\bm{q}\rightarrow 0$.
We find  
$\Gamma_{00}\propto\bra{\rm T}\Gamma\ket{\rm T}$ and 
$\Gamma_{11}\propto\bra{\rm S}\Gamma\ket{\rm S}$,  
where
$\ket{\rm T}=\ket{c\uparrow}\otimes\ket{c\uparrow}$ is an intraband triplet and 
$\ket{\rm S}=\frac{1}{\sqrt{2}}(\ket{c\uparrow}\otimes\ket{v\downarrow}-\ket{v\downarrow}\otimes\ket{c\uparrow})$ is an interband singlet. ($c$ and $v$ denote the conduction and valence bands, respectively.) 
Another intraband triplet, $\ket{v\downarrow}\otimes\ket{v\downarrow}$, corresponds to $\Gamma_{22}$.
The intraband singlet, $\frac{1}{\sqrt{2}}(\ket{c\uparrow}\otimes\ket{c\downarrow}-\ket{c\downarrow}\otimes\ket{c\uparrow})$, and intraband triplet, $\frac{1}{\sqrt{2}}(\ket{c\uparrow}\otimes\ket{c\downarrow}+\ket{c\downarrow}\otimes\ket{c\uparrow})$ do not appear in the present calculation.

With some straightforward calculations of eq. (\ref{kubo}), we obtain $\delta\sigma_{\rm W}$ as 
\begin{eqnarray}
\label{sigmaL}
&\delta\sigma_{\rm{W}}(L)=-\frac{e^2}{2\pi^2 \hbar}\nonumber\\
&\times\left(\alpha_{\rm t} {\rm log}\frac{\ell_{\rm t}^{-2}+\ell_{0}^{-2}}{\ell_{\rm t}^{-2}+L^{-2}}+\alpha_{\rm s} {\rm log}\frac{\ell_{\rm s}^{-2}+\ell_{0}^{-2}}{\ell_{\rm s}^{-2}+L^{-2}}\right),
\end{eqnarray}
where {\it L} is the size of the system. The first term of eq. (\ref{sigmaL}) corresponds to the contribution from the intraband triplet Cooperon, and the second term corresponds to that from the interband singlet Cooperon.
Under a magnetic field perpendicular to the two-dimensional lattice, we obtain the quantum correction to the magneto-conductivity in the following form
\begin{eqnarray}
\label{eq:MC}
&\delta\sigma_{\rm W}(B)=-\frac{e^2}{2\pi^2\hbar}\sum_{i={\rm s}, {\rm t}}\alpha_{i}\nonumber\\
&\times\left[\Psi\left(\frac{1}{2}+\frac{\ell_{B}^{2}}{\ell_{0}^2}+\frac{\ell_{B}^{2}}{\ell_{i}^2}\right)-\Psi\left(\frac{1}{2}+\frac{\ell_{B}^{2}}{\ell_{\phi}^2}+\frac{\ell_{B}^{2}}{\ell_{i}^2}\right)\right],
\end{eqnarray}
where $\Psi$ is the digamma function, $\ell_{\phi}$ is the coherent length, and $\ell_{B}=\sqrt{\hbar/4eB}$ is the magnetic length of the electron--electron pair.

Figure 2 shows the magnetic field dependence of $\delta\sigma_{\rm W}(B)$. A crossover from WL to WAL is obtained by increasing $E_F/\Delta$.
WL ($\delta \sigma >0$) arises when $E_F/\Delta\sim 1$, while WAL ($\delta \sigma <0$) arises when $E_F/\Delta$ exceeds a critical value of $E_c \simeq 3\Delta$. (At $E_F =E_c$, $\delta \sigma_{\rm W}\simeq 0$, because $\alpha_{\rm t} \simeq -\alpha_{\rm s}$ and $\ell_{\rm t} \simeq \ell_{\rm s}$.) Such a crossover can be also seen in the different context \cite{LuHaiZhou2011,ShanWenYu2012}.
In the zero field limit, eq. (\ref{eq:MC}) becomes
\begin{eqnarray}
\delta\sigma_{\rm W}(B\to0)\sim-\frac{e^2}{2\pi^2\hbar}\sum_{i={\rm s}, {\rm t}}\alpha_{i}{\rm log}\frac{\ell_{0}^{-2}+\ell_{i}^{-2}}{\ell_{\phi}^{-2}+\ell_{i}^{-2}},
\end{eqnarray}
which is equal to $\delta \sigma_{\rm W}(L\to \ell_{\phi})$.
\begin{figure}
 \begin{center}
  \includegraphics[width=8cm]{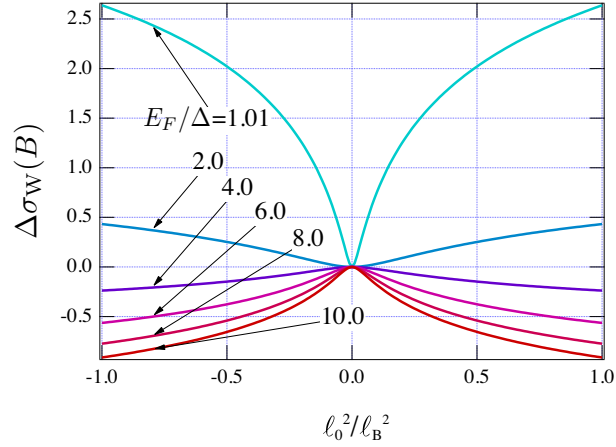}
 \end{center}
 \label{fig:MC}
 \caption{Magnetic field dependence of $\Delta\sigma_{\rm W}=[\delta\sigma_{\rm W}(B)-\delta\sigma_{\rm W}(0)]/(e^2/2\pi^2\hbar)$ for $\ell_{0}/\ell_{\phi}=$0.1. }
 \end{figure}
 
 \begin{figure}
 \begin{center}
  \includegraphics[width=8cm]{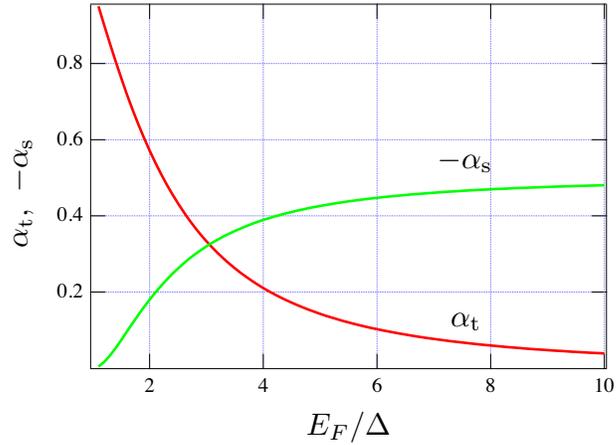}
 \end{center}
 \label{fig:alpha}
 \caption{Dependence of $\alpha_{\rm t}$ and $-\alpha_{\rm s}$ on $\lambda=E_F/\Delta$. }
 \end{figure}
 
The crossover from WL to WAL by changing $E_F$ can be understood in terms of the relationship between the intraband triplet ($\alpha_{\rm t}$) and the interband singlet ($\alpha_{\rm s}$) as follows. Figure 3 shows the $E_F$ dependence of the magnitude of $\alpha_{\rm t}$ and $-\alpha_{\rm s}$. For $E_F \lesssim E_c$, $\alpha_{\rm t}$ is greater than $-\alpha_{\rm s}$, i.e., the total quantum correction becomes positive, resulting in WL. However, for $E_F \gtrsim E_c$, $\alpha_{\rm s}$ is greater than $-\alpha_{\rm t}$, resulting in WAL. This property is clearly different from the conventional WAL of HLN theory in dilute SOC systems. 

\begin{figure}
 \begin{center}
  \includegraphics[width=8cm]{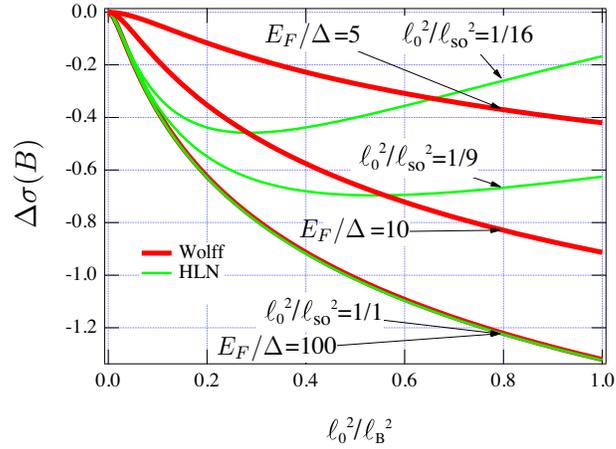}
 \end{center}
 \label{fig:MC}
 \caption{Dependence of $\Delta\sigma(B)=[\delta\sigma(B)-\delta\sigma(0)]/(e^2/2\pi^2\hbar)$ on the magnetic field in the WAL regime. 
The bold lines represent $\delta\sigma_{\rm W}$ and thin lines $\delta\sigma_{\rm HLN}$.
The parameter $\ell_{0}/\ell_{\phi}$ is common to $\delta\sigma_{\rm W}$ and $\delta\sigma_{\rm HLN}$. $\ell^{2}_{0}/\ell^{2}_{\phi}=$0.01. }
 \end{figure}

\section{Comparison of $\delta \sigma_{\rm W}$ and the conventional $\delta \sigma_{\rm HLN}$}
Fig. 4 compares the $\delta \sigma_{\rm W}$ in the WAL region and the conventional quantum correction obtained by using the HLN formula, $\delta \sigma_{\rm HLN}$. Here, $\delta \sigma_{\rm HLN}$ is calculated using the well-known HLN formula given by  
\begin{eqnarray}
\label{NormalHLN}
&\delta\sigma_{\rm HLN}(B)=-\frac{e^2}{2\pi^2\hbar}\Biggl[\alpha'_{\rm t}\Biggl\{\Psi\Bigg(\frac{1}{2}+\frac{\ell_B^2}{\ell_0^2}+\frac{\ell_B^2}{\ell_{\rm so}^2}+\frac{\ell_B^2}{\ell_{\phi}^2}\Biggr)\nonumber\\
&-\Psi\Bigg(\frac{1}{2}+\frac{4}{3}\frac{\ell_B^2}{\ell_{\rm so}^2}+\frac{\ell_B^2}{\ell_{\phi}^2}\Biggr)\Biggr\}\nonumber\\
&+\alpha'_{\rm s}\Biggl\{\Psi\Biggl(\frac{1}{2}+\frac{\ell_B^2}{\ell_{\phi}^2}+\frac{\ell_B^2}{\ell_{\rm so}^2}+\frac{\ell_B^2}{\ell_{\phi}^2}\Biggr)
-\Psi\Biggl(\frac{1}{2}+\frac{\ell_B^2}{\ell_{\phi}^2}\Biggr)\Biggr\}\Biggr].
\end{eqnarray}
Further, that the above formula is obtained by assuming the three-dimensional motion of electrons ($k_z \neq 0$) for the SOC scattering potential, although the kinetic energy is assumed to be two-dimensional. In this case, the coefficients of the intraband triplet and singlet are constant values, $\alpha'_{\rm t}=3/2$ and $\alpha'_{\rm s}=-1/2$. Therefore, the positive contribution of the triplet term becomes larger than the negative contribution of singlet term for sizable fields. Thus, the conventional HLN formula exhibits a minimum. However, for $\delta \sigma_{\rm W}$, the interband singlet term ($\alpha_{\rm s}$) always dominates the intraband triplet term ($\alpha_{\rm t}$) in the WAL region ($E_F>E_c$); therefore, $\delta \sigma_{\rm W}(B)$ monotonically decreases by the field, i.e., there is no minimum in $\delta \sigma_{\rm W}(B)$.
Strictly speaking, $\delta \sigma_{\rm W}(B)$ can exhibit a slight minimum when $2.92<E_F/\Delta <3.05$. However, the magnitude of $\delta \sigma_{\rm W}$ is almost zero in such a region because $E_F \simeq E_c$.

\section{WL in SOC lattice systems}
In the previous section, it was shown that the SOC lattice system exhibits both WL and WAL. It is rather surprising that WL arises in SOC lattices because WAL is expected in a case with SOC in general. In this section, we investigate the mechanism of WL in the SOC lattice. We obtain a better understanding of the mechanism of unconventional WAL by studying the mechanism of WL.

A remarkable WL appears when $E_F$ is close to the band edge, $E_F \simeq \Delta$. In such a region, the Foldy--Wouthuysen (FW) transformation \cite{FW1950} (the non-relativistic approximation) is useful for obtaining the low energy effective Hamiltonian, where the $4\times 4$ Wolff Hamiltonian is decoupled into the $2 \times 2$ conduction and valence band Hamiltonian. Through the FW transformation up to the order $(\gamma^2/\Delta)^2$, ${\cal H}_{\rm W}$ is approximately given by
\begin{eqnarray}
\label{FW}
{\cal H}'_{\rm W} &= \beta\Delta+\frac{\gamma^2{k}^2}{2\Delta}-\frac{\beta\gamma^4{k}^4}{8\Delta^3}+V({x})\nonumber\\
&+\frac{\gamma}{8\Delta^2}\nabla^2V(x)+\frac{\gamma^2}{4\Delta^2}\sigma\cdot[\nabla V(x)\times\bm{k}].
\end{eqnarray}
According to the Dirac theory, the third term is the kinetic energy correction, fifth term is the Darwin term, and sixth term is the SOC term. 
The Hamiltonian ${\cal H}_{\rm W}'$ is essentially equivalent to the conventional model by HLN (except for the third and fifth terms); therefore, we can calculate the quantum correction for ${\cal H}_{\rm W}'$ by the same procedure as the conventional HLN theory.
The relaxation time because of the SOC becomes 
\begin{eqnarray}
\label{HLNSO}
\frac{\hbar}{2\tau_{\rm so}}&=\pi\sum_{\bm{k}'}\braket{ \ |\bra{\bm{k},\beta}{\cal H}_{\rm so}\ket{\bm{k}',\alpha}|^2}_{\rm{imp}}\delta(\epsilon-E_{F})\nonumber\\
&=\pi n_{i}u_{0}^2\rho_{0}\frac{(\lambda-1)^2}{8}.
\end{eqnarray}
Here, we used $k_z=0$ for ${\cal H}_{\rm so}$, because we consider the two-dimensional Hamiltonian.

Now, the quantum correction for ${\cal H}_{\rm W}'$ is obtained in the following form:
\begin{eqnarray}
\label{2DHLN}
\delta\sigma_{\rm W}' (L)=-\frac{e^2}{2\pi^2\hbar}{\rm log}\frac{\ell_{\rm{so}}^{-2}+\ell_0^{-2}}{\ell_{\rm{so}}^{-2}+L^{-2}},
\end{eqnarray}
where the spin relaxation length is given by 
\begin{eqnarray}
\ell_{\rm{so}}^{-2}=\frac{\frac{2}{\tau_{\rm so}}}{D_{0}\left(1-\frac{2\tau}{\tau_{\rm so}}\right)}.
\end{eqnarray}
(Note that $\ell_{\rm so}$ is slightly different from the conventional definition of the spin-orbit relaxation-time, $\ell_{\rm so}=\sqrt{D_0\tau_{\rm so}}$.)
In this case, $\delta\sigma_{\rm W}'$ has only the triplet term \cite{HLN1980}, because $\tau_{\rm so}^x$ in the original HLN theory vanishes owing to $k_z=0$. Hence, $\delta \sigma_{\rm W}$ becomes positive and exhibits WL for $E_F < E_c$. 

Through the above analysis, we can clearly understand the characteristic of unconventional WAL for SOC lattice systems. In conventional dilute SOC systems, WAL arises only when we assume impurity scattering with spin relaxation (not spin-conserving) due to the three-dimensional motion through the impurity scattering. However, in SOC lattice systems, WAL arises even without the spin relaxation. The electrons travel through the conduction and valence bands, and their spins change owing to the SOC term $\bm{\sigma}\cdot \bm{k}$ in ${\cal H}_{\rm W}$. This interband SOC effect induces the interband singlet Cooperon instability, resulting in WAL only with spin-conserving impurity scattering.

\begin{figure}
 \begin{center}
  \includegraphics[width=8cm]{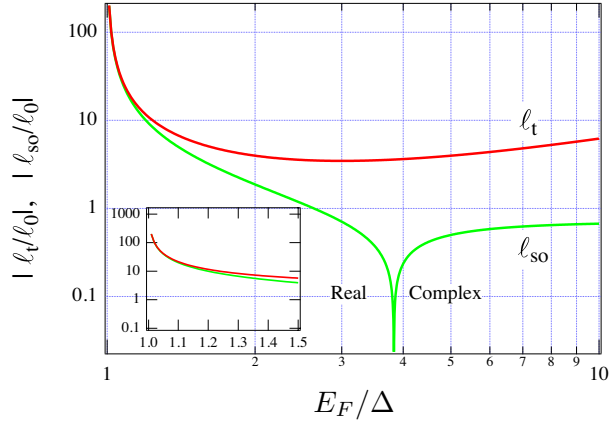}
 \end{center}
 \label{fig5}
 \caption{
 Dependence of spin relaxation length $\ell_{\rm t}$ and $\ell_{\rm so}$ normalized by $\ell_{0}$ on $\lambda=E_F/\Delta$
 }
\end{figure}

It is clear that $\delta \sigma_{\rm W}' (L)$ corresponds to the triplet part of eq. (\ref{sigmaL}), if we assume the correspondence $\ell_{\rm t} \leftrightarrow \ell_{\rm so}$. 
Figure 5 compares $\ell_{\rm so}$ and $\ell_{\rm t}$ as functions of $E_F$. When $E_F$ is close to the band-edge, i.e., in the WL region, $\ell_{\rm so}$ agrees with $\ell_{\rm t}$. If we expand $\ell^{-2}_{1}$ and $\ell^{-2}_{\rm so}$ around $E_F \sim \Delta$, we obtain 
\begin{eqnarray}
\frac{\ell^{-2}_{\rm t}}{\ell^{-2}_0}=\frac{1}{4}(\lambda-1)^2+{\cal O}((\lambda-1)^3),
\end{eqnarray}
\begin{eqnarray}
\frac{\ell^{-2}_{\rm so}}{\ell^{-2}_0}=\frac{1}{4}(\lambda-1)^2+{\cal O}((\lambda-1)^4).
\end{eqnarray}
Figure 5 also shows that $\ell_{\rm so}$ becomes an unphysical complex value for $E_F/\Delta >1+2\sqrt{2}$, whereas $\ell_{\rm t}$ is always a positive real value for the entire range of $E_F$. This is because $\ell_{\rm so}$ was obtained perturbatively with respect to $1/\tau_{\rm so}$.

\section{Discussions}
\begin{figure}
 \begin{center}
  \includegraphics[width=8cm]{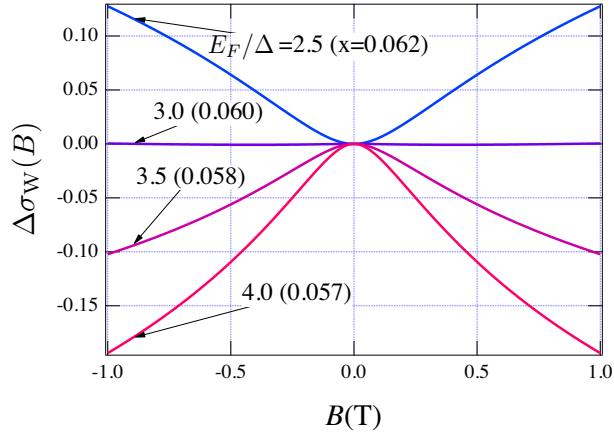}
 \end{center}
 \label{fig6}
 \caption{Dependence of $\Delta\sigma_{\rm W}(B)=[\delta\sigma_{\rm W}(B)-\delta\sigma_{\rm W}(0)]/(e^2/2\pi^2\hbar)$ on the magnetic field for  $E_F/\Delta$=2.5, 3.0, 3.5, and 4.0, when the Sb content of $\rm Bi_{1-\it x}Sb_{\it x}$ is {\it x}=0.062, 0.060, 0.058, and 0.057,
where $\ell_{0}=$10nm，$\ell_{\phi}=$200nm.}
 \end{figure}
Finally, we discuss the implications of the present theoretical results to the experiments. It is well known that Bi has a large SOC, and electrons at the $L$-point can be well described by the Wolff Hamiltonian \cite{Wolff1964,Fuseya2015}. The Fermi energy and band gap at the $L$-point of Bi is  $\Delta=7.7$ meV and $E_F=35.3$ meV, respectively \cite{SmithBaraffRowell1964,Fuseya2012,ZhuJPCM2018}. In this case, WAL is expected for pure Bi because $E_F/\Delta=4.6$ is greater than $E_c/\Delta \simeq 3$. In fact, WAL has been observed in Bi thin films \cite{Komori1983,Hirahara2014}. The WAL in Bi films has been interpreted by the dilute SOC scenario of HLN. However, the minimum of $\delta \sigma_{\rm HLN}(B)$, which is a characteristic property of the conventional HLN theory, has never been observed. Therefore, the monotonic decrease of $\delta \sigma (B)$ in Bi films can be interpreted by the present SOC lattice scenario.

Interestingly, $E_F$ of Bi can be changed by substituting Bi with Sb \cite{Wehrli1968,Fuseya2015}. The dependence of $E_F$ and $\Delta$ on the Sb content $x$ can be approximated as $E_F (x)/{\rm meV}=4.6-4.6x/0.09$ and $\pm\Delta(x) /{\rm meV}=1-x/0.04$ for Bi$_{1-x}$Sb$_{x}$ \cite{Fuseya2012}. It is expected that $E_F \simeq E_c$ at approximately $x_c \sim 0.06$. Therefore, the crossover from WAL to WL is expected in Bi$_{1-x}$Sb$_{x}$ through $x_c \sim 0.06$, as shown in Fig. 6. 

Studying the effect of pressure may be more suitable to see the WAL--WL crossover because this allows the localization effect by excluding the alloying. A good candidate for this would be PbTe, another typical Dirac electron system \cite{Dimmock1966, Hayasaka2016,Akiba2018, Izaki2019}. The band gap of PbTe can be reduced by applying pressure \cite{Papacon2002}. Because PbTe films at ambient pressure exhibit WAL \cite{Peres2014}, the crossover from WAL to WL is expected. However, note that a clear minimum in $\delta \sigma (B)$ is observed in PbTe \cite{Peres2014}, suggesting conventional WAL.

\section{Conclusion}
 We studied quantum correction to the conductivity in two-dimensional SOC lattice based on the Wolff Hamiltonian. We found that the interband singlet Cooperon due to the interband SOC effect results in WAL only with spin-conserving impurity scatterings. This is in contrast to conventional WAL in dilute SOC systems, where the intraband singlet Cooperon due to the spin-relaxation impurity scattering results in WAL.
The characteristics of this unconventional WAL in the SOC lattice are that $\delta \sigma_{\rm W} (B)$ never exhibits a clear minimum as a function of $B$, while a clear minimum in $\delta \sigma_{\rm HLN}(B)$ is observed for conventional WAL in dilute SOC system.
Furthermore, the crossover from WAL to WL is expected by the change in $E_F/\Delta$ in SOC lattice. This crossover can be observed in Bi$_{1-x}$Sb$_{x}$ or PbTe under pressure.

\section*{Acknowledgments}
We would like to thank M. Shiraishi and Y. Ando for their helpful discussions.
This work is supported by JSPS KAKENHI grants 19H01850 and 16K05437.

\section*{References}
\bibliographystyle{iopart-num}
\bibliography{refv4}

\end{document}